\documentclass[aps]{revtex4}

\usepackage{graphicx}

\newcommand{\beq}{\begin{equation}}
\newcommand{\enq}{\end{equation}}
\newcommand{\bea}{\begin{eqnarray}}
\newcommand{\ena}{\end{eqnarray}}
\begin{document}

\title{Construction of a giant vortex state in a trapped Fermi system}
\author{Emil Lundh}
\affiliation{Department of Physics, Ume{\aa} University, 
SE-901 87 Ume{\aa}, Sweden}

\begin{abstract}
A superfluid atomic Fermi system may support a 
giant vortex if the trapping potential is anharmonic. 
In such a potential, the 
single-particle spectrum has a positive curvature as a function of 
angular momentum. A tractable model 
is put up in which the lowest and next lowest Landau levels are 
occupied. Different parameter regimes are identified and characterized. 
Due to the 
dependence of the interaction on angular momentum quantum number, 
the Cooper pairing is at its strongest 
not only close to the Fermi level, but also close to the energy minimum. 
It is shown that the gas is superfluid in the interior of the 
toroidal density distribution and normal in the outer regions. 
Furthermore, the pairing may give rise to a localized density 
depression in configuration space. 
\end{abstract}
\maketitle

\section{Introduction}
Among the features that distinguish a superfluid from a normal 
fluid is the quantization of fluid circulation. This implies that 
a superfluid responds to external rotation by forming one or more 
vortices, each of which carries an integer number of circulation 
quanta. In an infinite homogeneous fluid, it is well known that 
singly quantized vortices are energetically favorable for a given 
rotation rate, since the angular momentum is proportional to the 
quantum number and the energy is proportional to its square
\cite{ll,pethick2001}. As a result, a rotating homogeneous 
superfluid does normally not form giant vortices, i.~e., vortices 
with quantum number larger than one, but instead a lattice of 
singly quantized vortices will form. Such 
lattices are also observed in bosonic \cite{madison2000,ketterle2001} and 
fermionic \cite{ketterle2005} superfluid gases contained in 
harmonic trapping potentials. The parameters in these 
experiments are such that the system is much larger than the core 
of a single vortex, so that the energy minimization argument 
sketched above goes through. 

The situation is different if 
the length scale of a vortex is comparable to the system radius. 
In finite superconductors, it has been found numerically that 
giant vortices 
form in small samples at strong fields \cite{deo1997}. 
For a Bose condensate, the quantum number of a vortex depends on the 
shape of the potential used to confine the atoms as well as on 
parameters associated with rotation and inter-particle interactions. 
When the potential in the radial direction is harmonic, an array of 
singly quantized vortices is the energetically favorable state for 
all parameter values \cite{butts1999,kavoulakis2000}, but in an anharmonic 
potential, a Bose condensate will develop a giant vortex 
if the rotation speed is 
high enough and the interaction energy is weak enough 
\cite{lundh2002,kasamatsu2002,fischer2003,kavoulakis2003,aftalion2003,
jackson2004,fetter2005,fu2006,bargi2006,lundh2006}. 

Because a condensed Bose gas at zero temperature obeys a nonlinear 
Schr{\"o}dinger equation that in the noninteracting limit reduces to 
the single-particle Schr{\"o}dinger equation 
\cite{pethick2001,kavoulakis2000}, it 
is possible to put up an exact argument proving the energetic stability 
of giant vortices in anharmonic traps and their instability in harmonic 
traps \cite{lundh2002}. In a Fermi gas, no such criteria have yet 
been established. Therefore, as a first step towards 
understanding the 
rotational properties in superfluid Fermi gases, this paper shows 
that it is possible, within a BCS ansatz, to construct a 
situation where a giant vortex is the energetically favorable state. 
In order to make the problem tractable, the study is confined to the 
so-called lowest Landau level (LLL), i.~e., the space of states 
without radial excitations; the extension to the next lowest Landau 
level will also be studied. 
In Sec.\ \ref{sec:lll} the physical setting is described and definitions
are introduced. In Sec.\ \ref{sec:bdg} the solution of the equations is 
carried out. Section \ref{sec:params} discusses the physics of the solution 
in different parameter regimes. In Sec.\ \ref{sec:nlll} the effects of 
extending the space of occupied states and approaching more realistic 
parameter values are discussed. 
Finally, Sec.\ \ref{sec:conclusions} 
provides a conclusion and outlook.

\section{Lowest Landau level}
\label{sec:lll}
\begin{figure}
\includegraphics[width=\columnwidth]{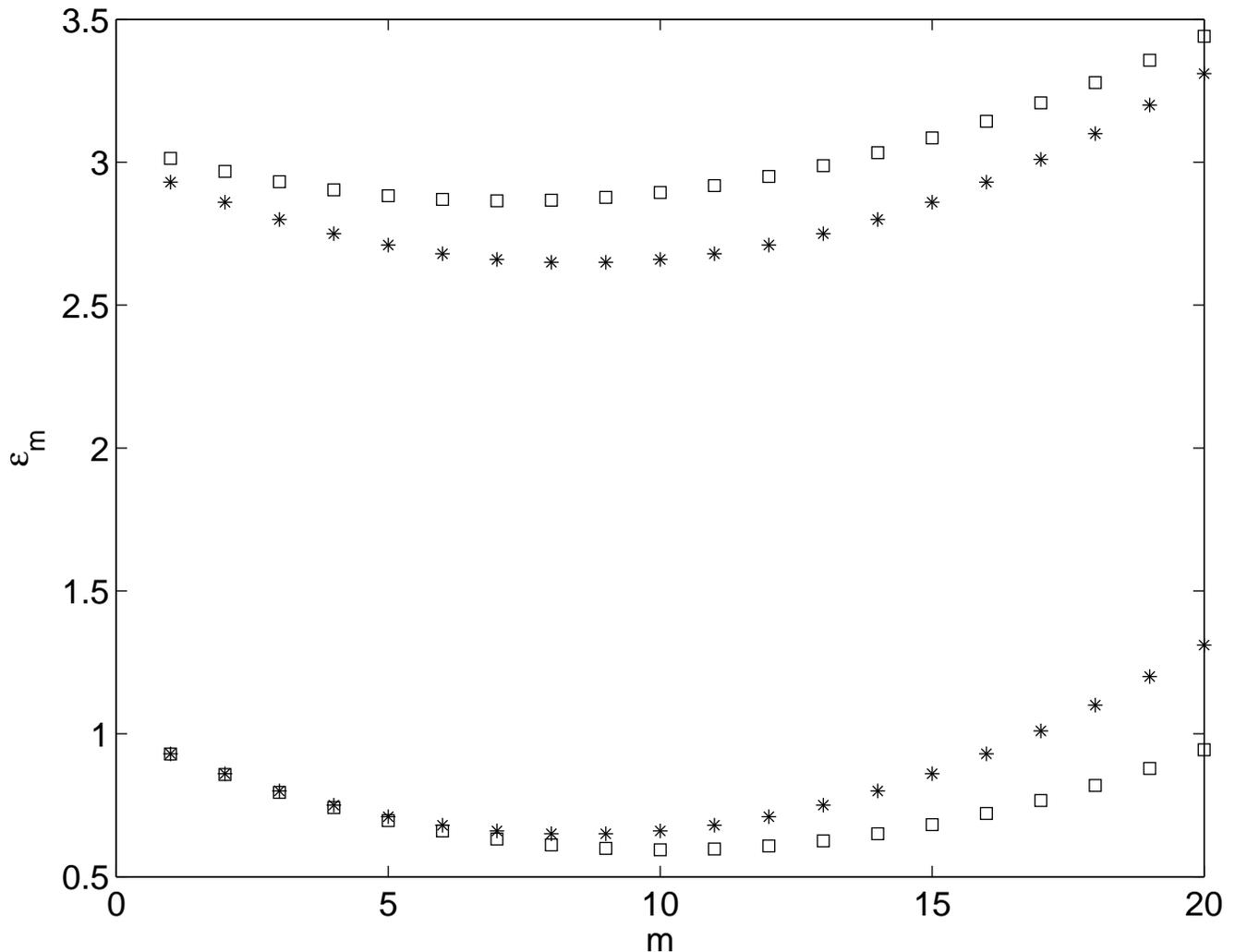}
\caption{\label{fig:levels}
Energy levels in the two lowest Landau levels in an anharmonic trap with 
external rotation frequency $\Omega = 1.1$ and 
anharmonicity $a=0.01$. Squares are the exact numerically 
computed energy levels. Asterisks denote the variational 
approximation using harmonic-oscillator eigenfunctions.}
\end{figure}

Consider a system of spin-$1/2$ fermions confined in an 
anharmonic, i.~e.\ quadratic plus quartic, trapping potential. The fermions
are subject to a rotational force 
with a rotation frequency $\Omega$. (Alternatively, $\Omega$ can 
be seen as a Lagrange multiplier applied to ensure a finite 
angular momentum $L_z$). The number of particles in each 
spin state is $N$, so that the total number is $2N$. The 
confinement in the direction along the rotation axis is 
assumed to be so tight 
that the motion in that direction is frozen out, so that the 
system can be assumed to be two-dimensional. 
The Hamiltonian is
\beq
\label{storhamilton}
H = \sum_{\sigma}\int dr \psi_{\sigma}^{\dagger}(r) H_0
\psi_{\sigma}(r) + 
g \int dr \psi_{\uparrow}^{\dagger}(r)
\psi_{\downarrow}^{\dagger}(r) \psi_{\downarrow}(r) 
\psi_{\uparrow}(r),
\enq
where $\sigma$ runs over the two spin indices $\uparrow,\downarrow$, 
and the interaction strength $g$ is assumed negative, $g=-|g|$.
Choosing units 
so that the particle mass and Planck's constant $\hbar$ are unity, 
the single-particle Hamiltonian is 
\beq
H_0 = -\frac12 \nabla^2 + V(r) -\Omega \hat{L}_z.
\label{hamiltonian}
\enq
The potential is
\beq
V(r) = \frac{\omega^2}{2} r^2(1+a r^2),
\enq
so that $\omega$ is the trap frequency associated with the harmonic 
part and $a$ is the degree of anharmonicity. The passage 
to dimensionless units is completed by setting $\omega=1$. 
The single-particle spectrum associated with 
the Hamiltonian (\ref{hamiltonian}) can be solved numerically, 
resulting in a positive curvature for the energy as a function 
of angular quantum number $m$ for a fixed radial quantum number 
$n_r$, as seen in Fig.\ \ref{fig:levels}.
In order to make the problem tractable one needs to assume 
that no states are occupied except those in the lowest 
Landau level (LLL), i.~e.\ the levels with no radial nodes, $n_r=0$. 
The restrictions on physical parameters needed for this assumption 
to hold will shortly be explored. 
In general, the single-particle spectrum in the LLL around the minimum can be 
approximated with a quadratic expression,
\beq
\epsilon_{\kappa} = \epsilon_0 + \frac{\alpha}{2} {\kappa}^2,
\enq
where $\kappa=m-m_0$ is a shifted azimuthal quantum number; the 
offset $m_0$ is the quantum number at which the 
energy has its minimum; and the curvature $\alpha$ has to, in 
general, be computed numerically.
When the anharmonicity $a$ is very small, the energy spectrum can 
be solved perturbatively using harmonic-oscillator 
eigenfunctions, resulting in
\beq
\epsilon_m = \omega -(\Omega-1)m + 
\frac{a}{2}(m+1)(m+2).
\enq
Thus, in this limit the minimum-energy quantum number is given by 
\beq
\label{defm0}
m_0= (\Omega-1)/a-3/2,
\enq
the curvature is $\alpha=a$, and 
the ground-state energy $\epsilon_0 = 1-m_0^2+a$. For not so 
small $a$, or high enough quantum numbers $m$, the corrections 
to the perturbative expressions become important, but they only 
result in quantitative shifts. In the following, we shall use 
the effective curvature parameter $\alpha$, which may be slightly 
different from $a$, in the calculations, but the more physical 
discussions and order-of-magnitude estimates 
will be stated in terms of the actual anharmonicity $a$.

The number of particles is determined by the chemical potential 
$\mu$, which in the weakly interacting limit is equal to the 
Fermi energy. For convenience, a Fermi quantum number 
$\kappa_F$ is defined through the relation
\beq
\mu = \epsilon_0 + \frac{\alpha}{2}\kappa_F^2,
\enq
so that the actual number of particles per spin state is $N=2\kappa_F$ 
in the weakly interacting limit, but one must bear in mind that it 
may differ from that when interactions are finite. 
The parameters are assumed to be such that $\mu < 0$, implying 
$m_0-\kappa_F>0$. That way, 
the occupied part of the spectrum is bounded from below at some 
finite angular momentum quantum number $m>0$, and one can treat the 
spectrum as if it is extended indefinitely in both the positive 
and negative $\kappa$-direction. 

The coupling constant $g$ introduced in Eq.\ (\ref{storhamilton}) 
is a pseudopotential whose 
relation to the actual scattering length $a_s$ is in 
general nontrivial 
\cite{sensarma2006}, 
but in the weak-coupling limit it reduces to the simple 
form $g=4\pi a_s$.
The conventional dimensionless coupling parameter in the study of 
trapped Fermi gases is the product $\zeta=k_Fa_s$, the 
Fermi wavenumber times
the s-wave scattering length. In terms of 
our Fermi quantum number $\kappa_F$ and coupling constant $g$, we 
obtain $\zeta = \alpha^{1/2}\kappa_Fg/4\pi$, and we shall confine the 
study to small values of $\zeta$, i.~e. weak coupling, 
in order not to violate the 
conditions for the model to be valid. 
We now discuss the conditions for the LLL assumption to hold. 
Firstly, the next lowest Landau level, with $n_r=1$, must lie 
higher than the chemical potential. The spacing between the Landau levels 
is $2$ (in the present units where $\omega=1$), so the condition is 
$\mu < \epsilon_0+2$ or in terms of physical quantitites,
\beq
\label{ineq1}
\frac{a}{2}\left(\frac{N}{2}\right)^2 < 2.
\enq
Furthermore the condition $\mu<0$ stated above needs to be improved. 
Interactions broaden the occupation of fermions and smear out the 
Fermi level on a scale $|g|$, the interaction strength. 
The distance between the lower Fermi level and the edge of the 
spectrum at $m=0$ thus needs to be larger than $|g|$, that is, 
$m_0-\kappa_F > |g|$ or
\beq
\label{ineq2}
\frac{\Omega-1}{a} - \frac{N}{2} > |g|.
\enq
Note that according to the first inequality, Eq.\ (\ref{ineq1}), 
$a$ must be 
of order $N^{-2}$ or smaller which means that the second restriction 
(\ref{ineq2}) is 
automatically fulfilled as long as $\Omega-1$ is positive 
and not smaller than $|g|N^{-2}$, which is an extremely small quantity. 
The first inequality does, however, put quite severe limitations on the 
experimental 
implementation of the state, since it forces the anharmonicity to be 
less than or of the order of the inverse square of the number of 
particles. These matters will be discussed further in the concluding 
section.

\section{Bogoliubov-de Gennes analysis}
\label{sec:bdg}
Through a number of assumptions on the physical parameters we have 
arrived at a description of a Fermi gas with a quadratic 
single-particle spectrum. This is the textbook case, and in principle, 
one may simply apply the usual textbook BCS expressions known for 
superconductors \cite{ll}. 
Two things remain to be sorted out: first, the gap must be assumed to 
be larger than the level spacing, which permits us to treat the 
levels as a continuum (for paired Fermi systems with a 
discrete spectrum, see Ref.\ \cite{delft}). 
Second, one must consider the variation of the 
coupling with quantum number $\kappa$. This will have an effect on the 
density profile and quasi-particle energy spectrum, as we shall see.

The Bogoliubov-de Gennes (BdG) equation in two dimensions reads 
\beq
\left(\begin{array}{ll}
H_0-\mu & \Delta(r,\theta) \\
\Delta(r,\theta)^* & -H_0^*+\mu 
\end{array}\right)
\left(\begin{array}{l}
u_{\kappa}(r,\theta) \\
v_{\kappa}(r,\theta) 
\end{array}\right) 
= E_{\kappa}
\left(\begin{array}{l}
u_{\kappa}(r,\theta) \\
v_{\kappa}(r,\theta) 
\end{array}\right),
\label{bdg_cyl}
\enq
where we have anticipated that the solutions $u_{\kappa},v_{\kappa}$ can be 
labeled by the shifted angular momentum quantum number ${\kappa}$.
When the single-particle energy spectrum is symmetric around 
$m=m_0$ (or ${\kappa}=0$), the gap function can be written
\beq
\Delta(r,\theta) = \Delta(r) e^{-i 2m_0 \theta}.
\enq
This amounts to assuming a cylindrically symmetric gap function 
and corresponds to the usual assumption of a homogeneous gap 
function in an infinite system. 
In order to match the $\theta$ dependence, the amplitudes take on 
the form
\bea
u_{\kappa}(r) = u_{\kappa} \phi_{m_0+{\kappa}}(r),\nonumber\\
v_{\kappa}(r) = v_{\kappa} \phi_{m_0-{\kappa}}^*(r),
\ena
where $\phi_{m}$ are single-particle eigenfunctions in the 
anharmonic trap and $u_{\kappa}$ and $v_{\kappa}$ are space-independent 
amplitudes. 
The BdG equation (\ref{bdg_cyl}) is now 
multiplied through by eigenfunctions $\phi_{m}$ and 
integrated, resulting in 
\beq
\left(\begin{array}{cc}
\xi_{\kappa} & \Delta_{\kappa} \\
\Delta_{\kappa}^* & -\xi_{\kappa} 
\end{array}\right)
\left(\begin{array}{l}
u_{\kappa} \\
v_{\kappa} 
\end{array}\right) 
= E_{\kappa}
\left(\begin{array}{l}
u_{\kappa} \\
v_{\kappa} 
\end{array}\right),
\enq
where 
\beq
\label{cmdef}
\Delta_{\kappa} = 
\int dr 2\pi r \Delta(r) \phi_{m_0+\kappa}^*(r) \phi_{m_0-\kappa}^*(r),
\enq
and 
\beq
\xi_{\kappa} = \varepsilon_{\kappa} - \mu = \frac{\alpha}{2}({\kappa}^2-{\kappa}_F^2).
\enq
The solution for the energies and amplitudes is similar to the
traditional BCS expressions:
\bea
E_{\kappa} = \sqrt{\xi_{\kappa}^2 + |\Delta_{\kappa}|^2},
\nonumber\\
u_{\kappa}^2 = \frac12(1+\frac{\xi_{\kappa}}{E_{\kappa}}),
v_{\kappa}^2 = \frac12(1-\frac{\xi_{\kappa}}{E_{\kappa}}).
\label{lllsolution}
\ena

Now turn to the self-consistency equation for the gap function,
\beq
\Delta(r) = |g|\sum_{\kappa} u_{\kappa}(r) v^*_{\kappa}(r). 
\enq
Inserting the calculated Bogoliubov amplitudes into the 
self-consistency equation, one obtains
\bea
\label{selfconsistency}
\Delta(r) 
&=& |g|\sum_{\kappa} \phi_{m_0+{\kappa}}(r) \phi_{m_0-{\kappa}}(r)
\frac{|\Delta_{\kappa}|}{2E_{\kappa}}.
\ena
Multiply by $\phi_{m_0+\kappa'}(r) \phi_{m_0-\kappa'}(r)$, integrate and 
rename the subscripts. The result is 
\beq
\label{selfcons2}
\Delta_{\kappa} = |g| \sum_{\kappa'} \frac{|\Delta_{\kappa'}|}{2E_{\kappa'}}
V_{\kappa,\kappa'}
\enq
where the coupling matrix element was defined  as
\bea
V_{\kappa,\kappa'} &=& \langle m_0+\kappa,m_0-\kappa | 
m_0+\kappa',m_0-\kappa'\rangle {\rm , and}
\nonumber\\
\langle m,n | k,l\rangle &=& \int dr 
\phi_m^* \phi_n \phi_k \phi^*_l. 
\ena
In order to be able to produce a definite expression, let us make use of 
the results for a harmonic trap. Since Eq.\ (\ref{ineq1}) confines the 
parameters to lie in the regime of a very weak anharmonicity, 
the anharmonic trap 
can at most bring in small quantitative corrections. The matrix element 
for the contact potential in the space of harmonic-oscillator eigenfunctions is
\cite{kavoulakis2000}
\beq
\langle m,n | k,l\rangle
= \frac{1}{2\pi} \frac{[(|m|+|n|+|k|+|l|)/2]!}{2^{(|m|+|n|+|k|+|l|)/2}
\sqrt{|m|! |n|! |k|! |l|!}}.
\enq
Note that the potential is separable:
\beq
V_{\kappa,\kappa'} \equiv V_{\kappa}V_{{\kappa}'}, \, 
V_{\kappa} = \sqrt{\frac{(2m_0)!}{2\pi 2^{2m_0}(m_0+\kappa)!(m_0-\kappa)!}},
\enq
and furthermore, the potential is in the limit of large $m_0$ 
approximated by
\beq
\label{vapprox}
V_{\kappa} 
\approx \frac{1}{\sqrt{2\pi}(\pi m_0)^{1/4}}e^{\frac{-\kappa^2}{2m_0}}.
\enq
A separable potential presents a considerable simplification. 
The solution for the gap is
\beq
\Delta_{\kappa} = \Delta V_{\kappa},
\enq
where the constant $\Delta$ is found by solving the self-consistency
equation in its final form,
\beq
\label{finalself}
1 = |g|\sum_{\kappa} \frac{V_{\kappa}^2}{\sqrt{\Delta^2V_{\kappa}^2+\xi_{\kappa}^2}}.
\enq
This sum has, in general, to be computed numerically. Nevertheless, 
there is plenty of information to be extracted from its qualitative 
form, as we shall explore in the next section.

\section{Parameter regimes}
\label{sec:params}
Consider the self-consistency equation (\ref{finalself}).
If the coupling is weak, $\Delta$ is expected to be small as well. 
In this case the largest contribution to the self-consistency sum 
comes from the states close to the Fermi surface, and 
the usual textbook BCS solution is recovered \cite{ll},
\beq
\Delta = \omega_c e^{-1/(N(0)V_{00}|g|)},
\enq 
where $\omega_c$ is the cut-off frequency for the sum and 
$N(0)$ is the density of states at the Fermi level.
However, that solution is only valid when the interaction strength 
at the Fermi surface is larger than the separation of the energy 
levels; if it is smaller, one cannot neglect discreteness effects 
\cite{delft}. This criterion translates to
$
|g| e^{-\kappa_F^2/m_0}/\sqrt{\pi m_0} > \alpha \kappa_F.
$
On the other hand, if $|g|$ is too large, pairing will happen 
away from the Fermi surface, and as will shortly be shown, this occurs when 
$|g| > \alpha \kappa_F^2$. Combining the two inequalities, one finds 
that a criterion for avoiding {\it both} discreteness {\it and} 
off-Fermi surface pairing is 
\beq
\frac{\kappa_F}{\sqrt{\pi m_0}} > e^{\kappa_F^2/m_0},
\enq
which cannot be fulfilled for any value of $\kappa_F$. 
We conclude that the standard textbook solution 
is not applicable in the present system.

\begin{figure}
\includegraphics[width=0.48\textwidth]{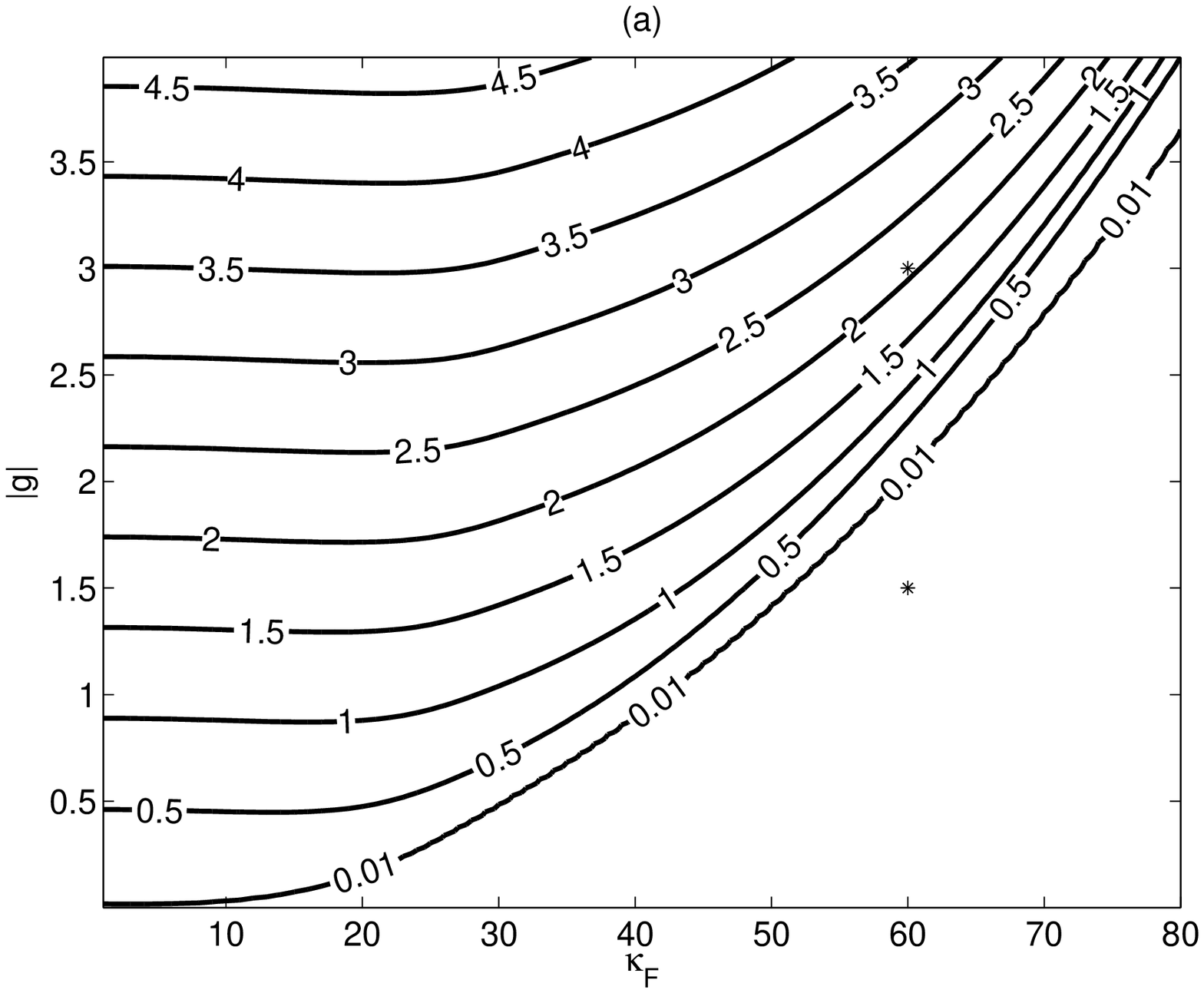}
\includegraphics[width=0.48\textwidth]{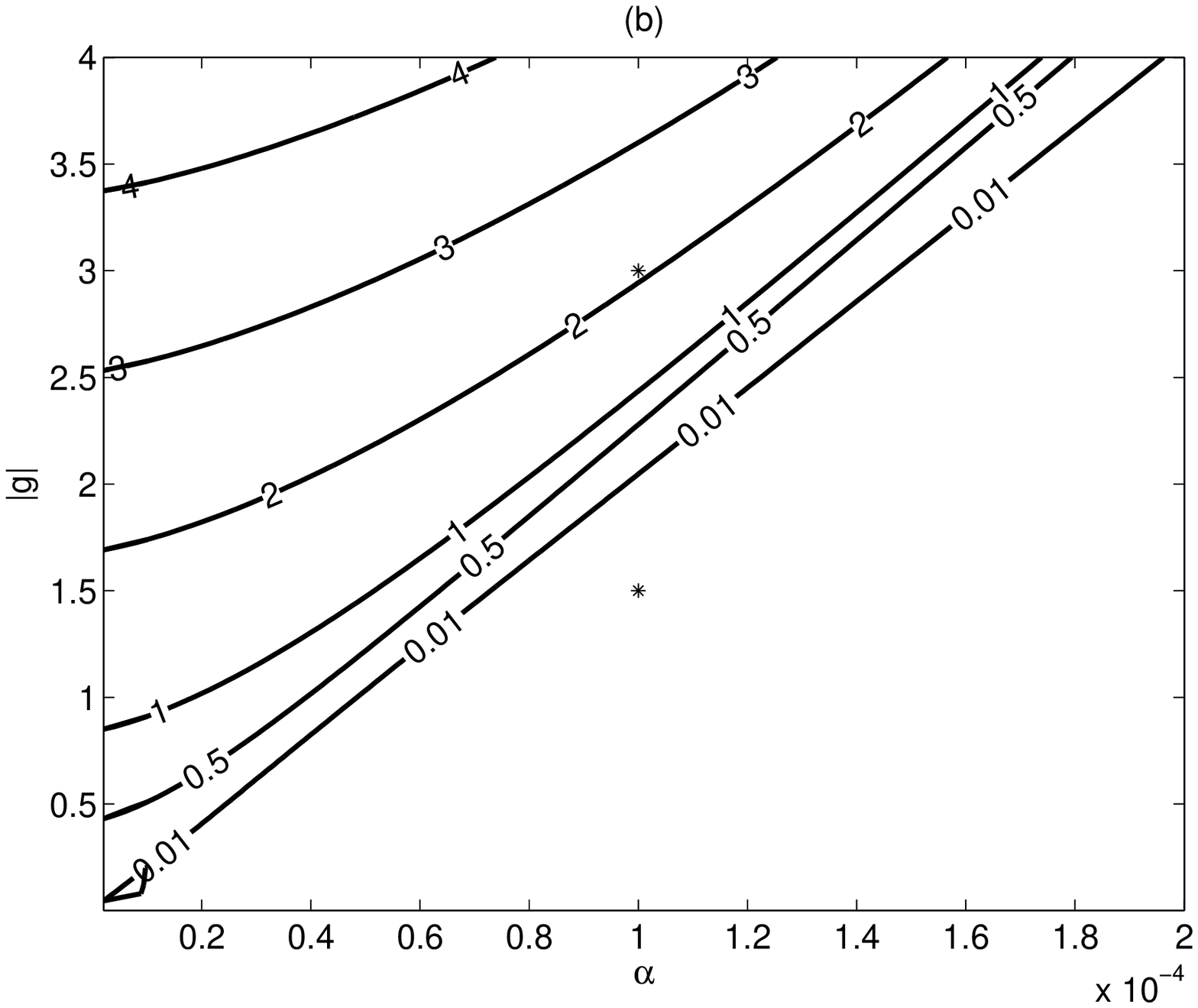}
\includegraphics[width=0.48\textwidth]{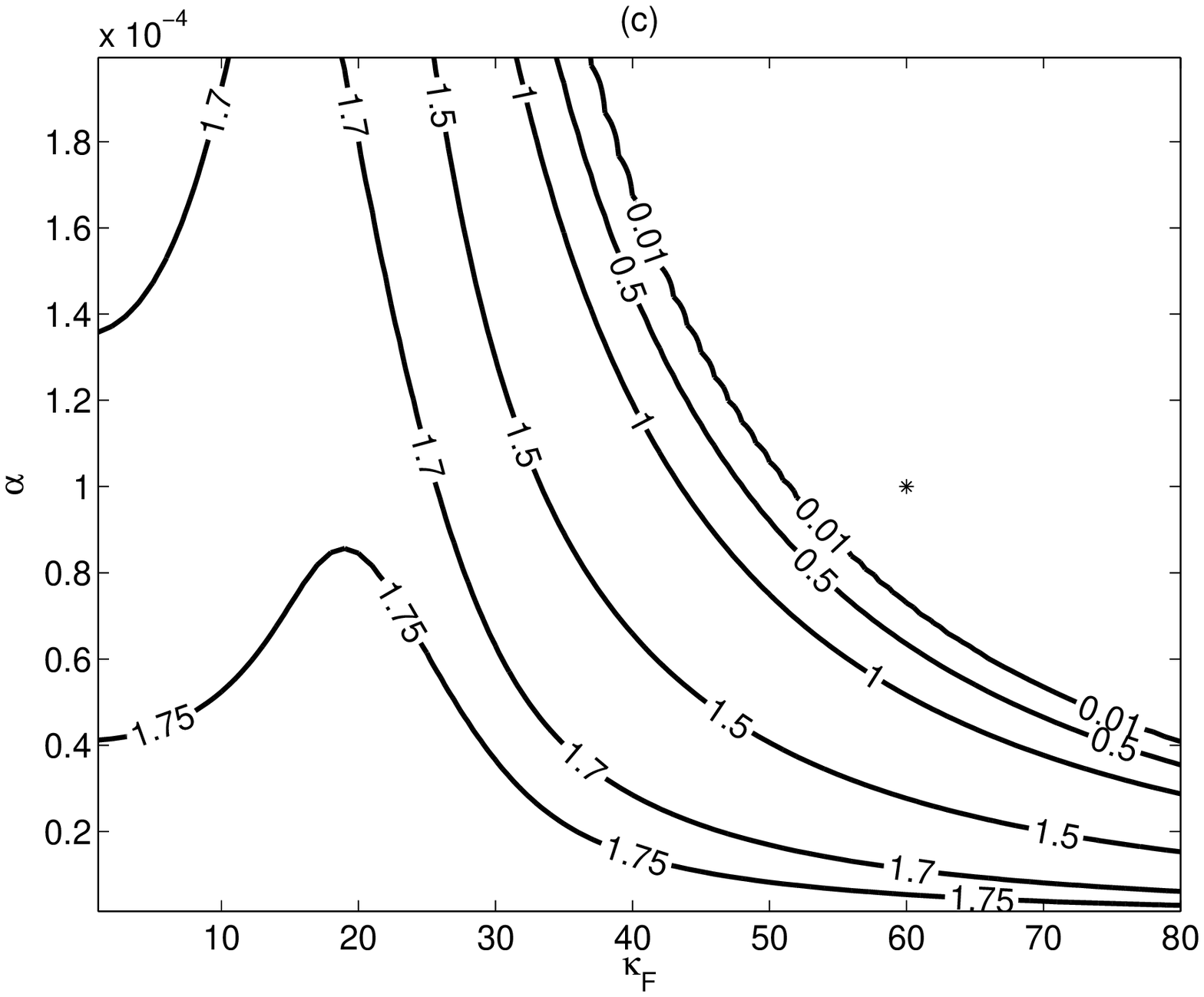}
\caption{\label{fig:grovphase}
Level curves for the gap function $\Delta$ in three slices 
of parameter space. (a) Fixed anharmonicity $\alpha=0.0001$, 
(b) fixed Fermi quantum number $\kappa_F=60$,
(c) fixed coupling $|g|=1.5$. The minimum-energy quantum number 
is fixed to $m_0=100$ in all three panels. Asterisks denote the 
phase-space points used in Fig.\ \ref{fig:qpspectrum}.
}
\end{figure}

We now derive the criterion already used above, for 
states away from the Fermi surface to contribute to the 
self-consistency sum. When the sum is exhausted by terms far 
from the Fermi surface we can write 
\beq
|g| \sum_{|\kappa| \ll \kappa_F} \frac{V_{\kappa}^2}{\sqrt{\Delta^2 V_{\kappa}^2 + \xi_{\kappa}^2}} = 1.
\enq
Setting $\Delta=0$ yields an inequality, 
\beq
|g|\sum_{|\kappa| \ll \kappa_F} \frac{V_{\kappa}}{|\xi_{\kappa}|} > 1.
\enq
The sum is estimated by inserting the maximum value of the 
summand and multiplying it by the effective range of the 
potential $V_{\kappa}$,
\beq
|g| \sqrt{2 m_0} \frac{V_0^2}{|\xi_0|} > 1,
\enq
where $\sqrt{2m_0}$ is the width of the function $V_{\kappa}$ and 
$V_0^2=  (4\pi^3 m_0)^{-1/2}$ 
is its value in the region 
where the summand is at its largest. The factors involving 
$m_0$ cancel out, so the critical coupling 
strength for an off-Fermi surface contribution reduces to
\beq
\label{condition}
|g| > g_c \propto |\xi_0| \propto \alpha \kappa_F^2. 
\enq
In Fig.\ \ref{fig:grovphase}, 
the numerically computed value of $\Delta$ is plotted in various 
slices of parameter space. It is seen how the $\Delta=0.01$ level curve 
can be reasonably well fitted to Eq.\ (\ref{condition}) in all 
three plots. Beyond this curve, the discreteness of the 
spectrum is expected to be important and the BCS solution cannot be 
trusted to be quantitatively correct.

\begin{figure}
\includegraphics[width=0.48\columnwidth]{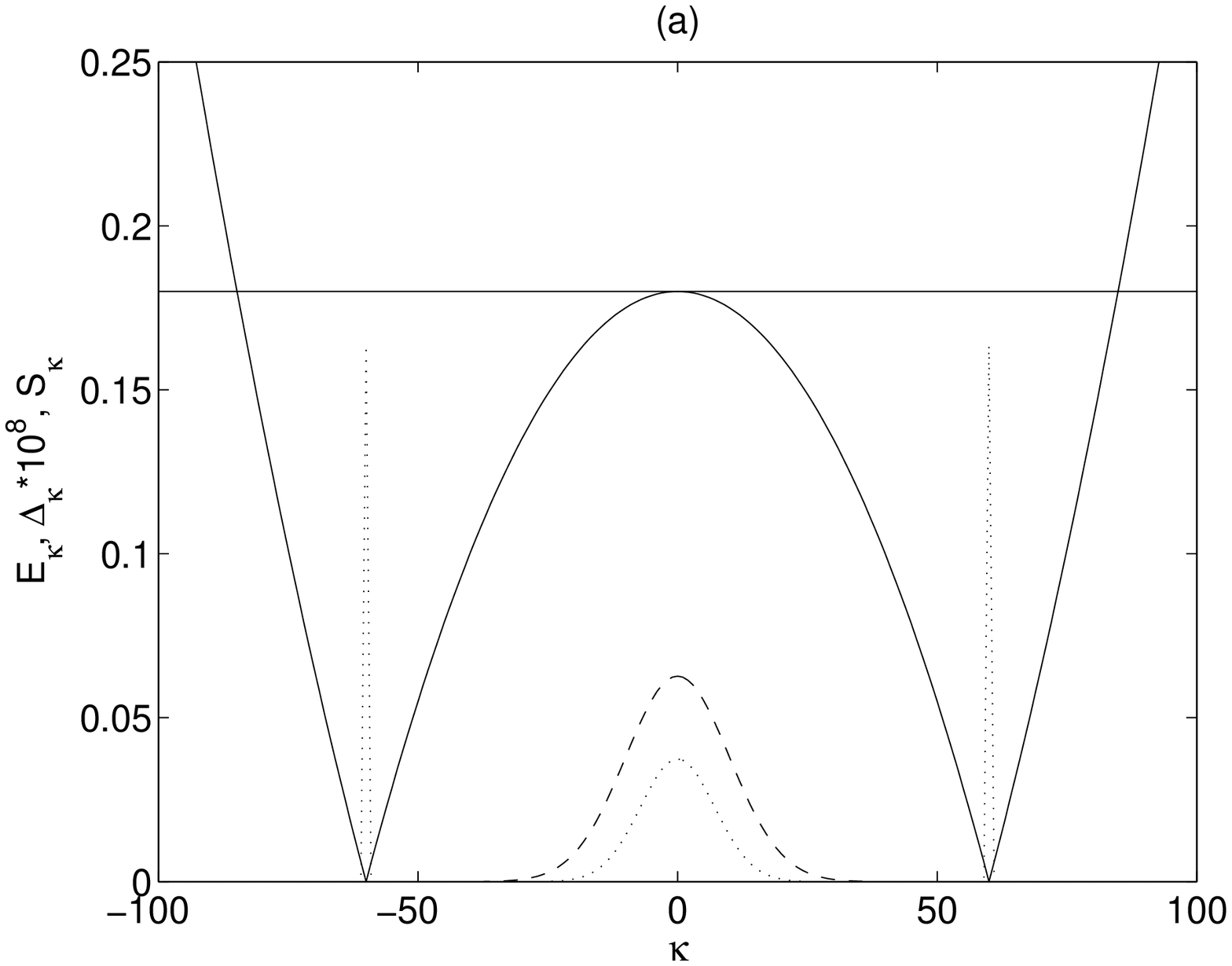}
\includegraphics[width=0.48\columnwidth]{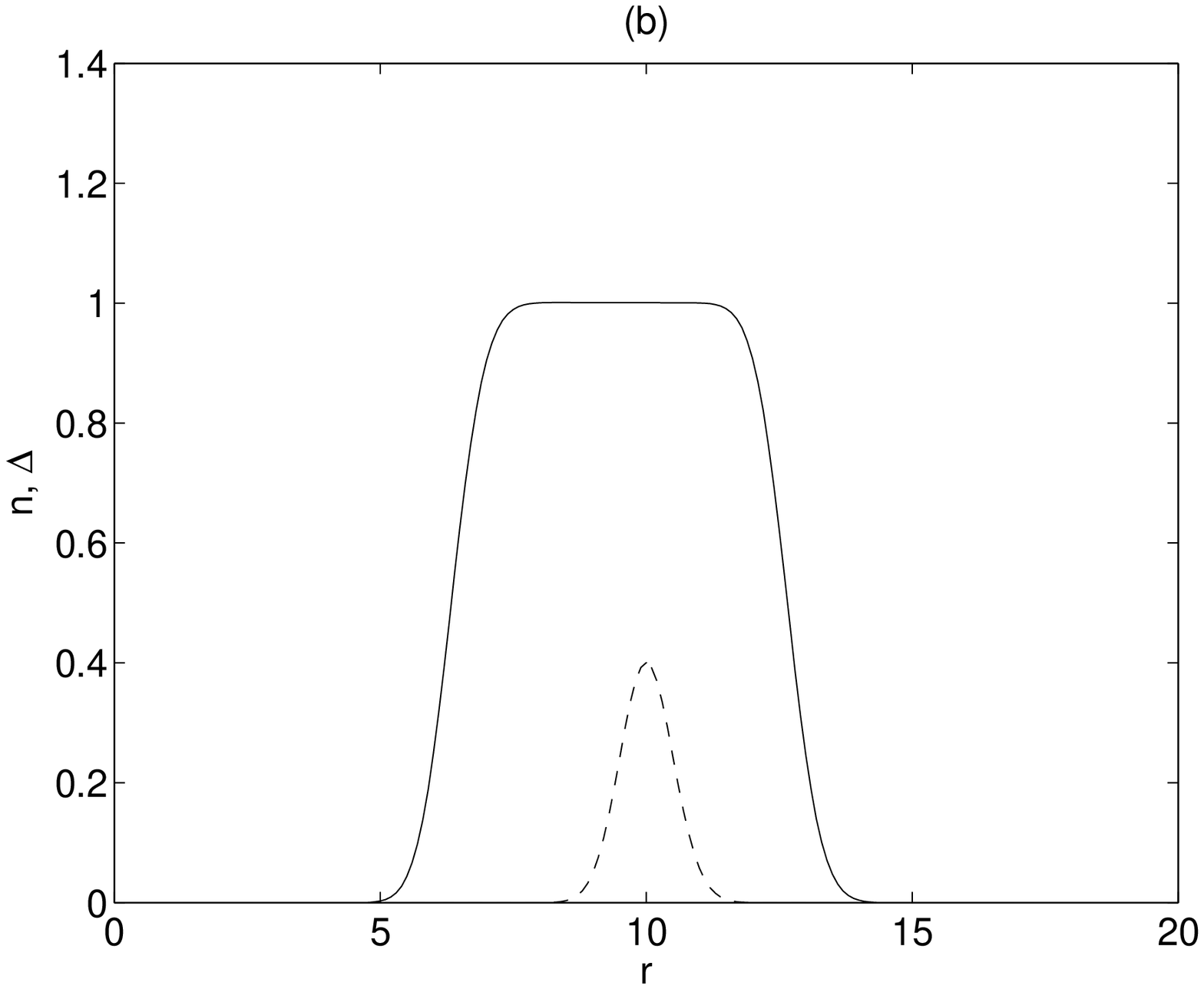}
\includegraphics[width=0.48\columnwidth]{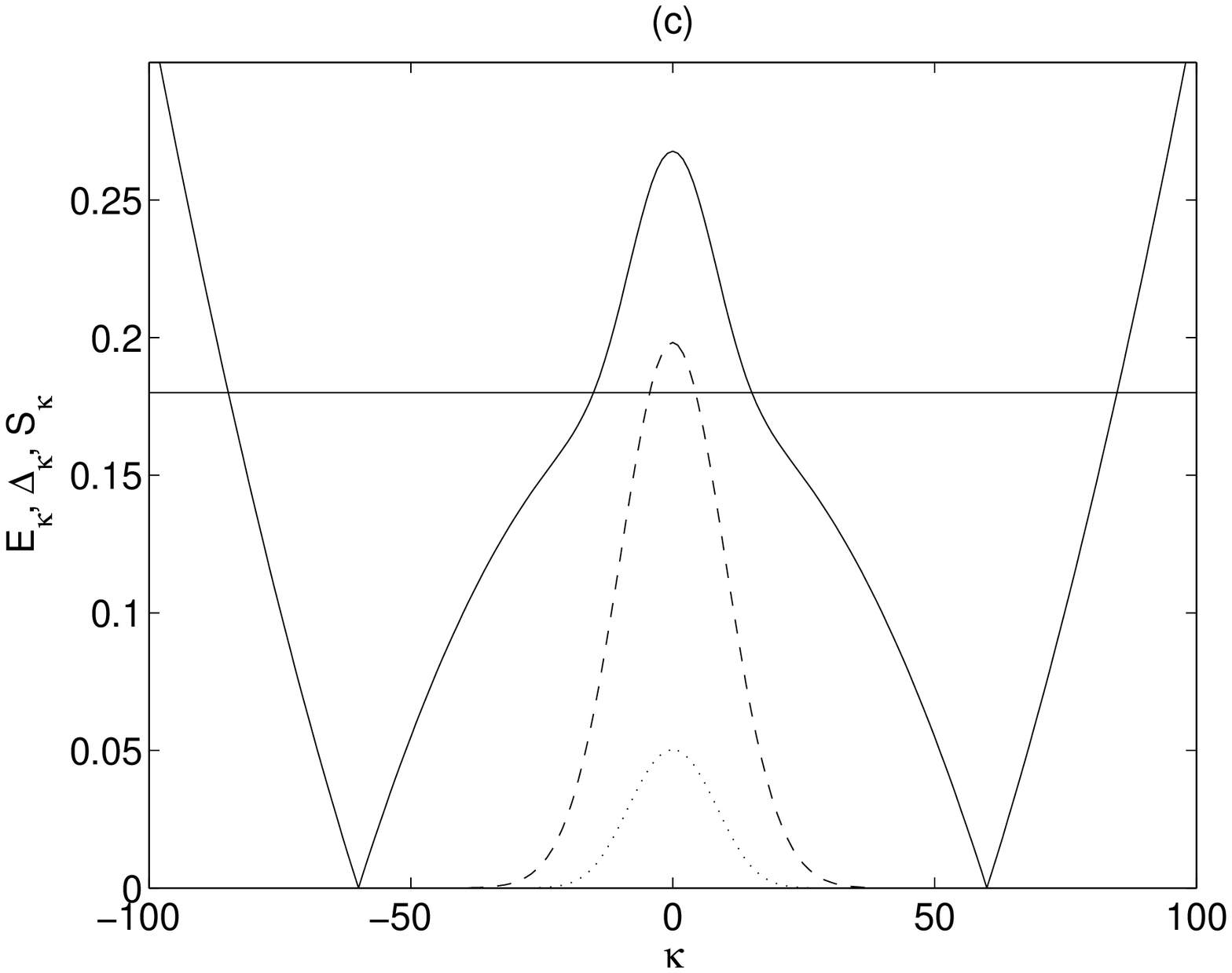}
\includegraphics[width=0.48\columnwidth]{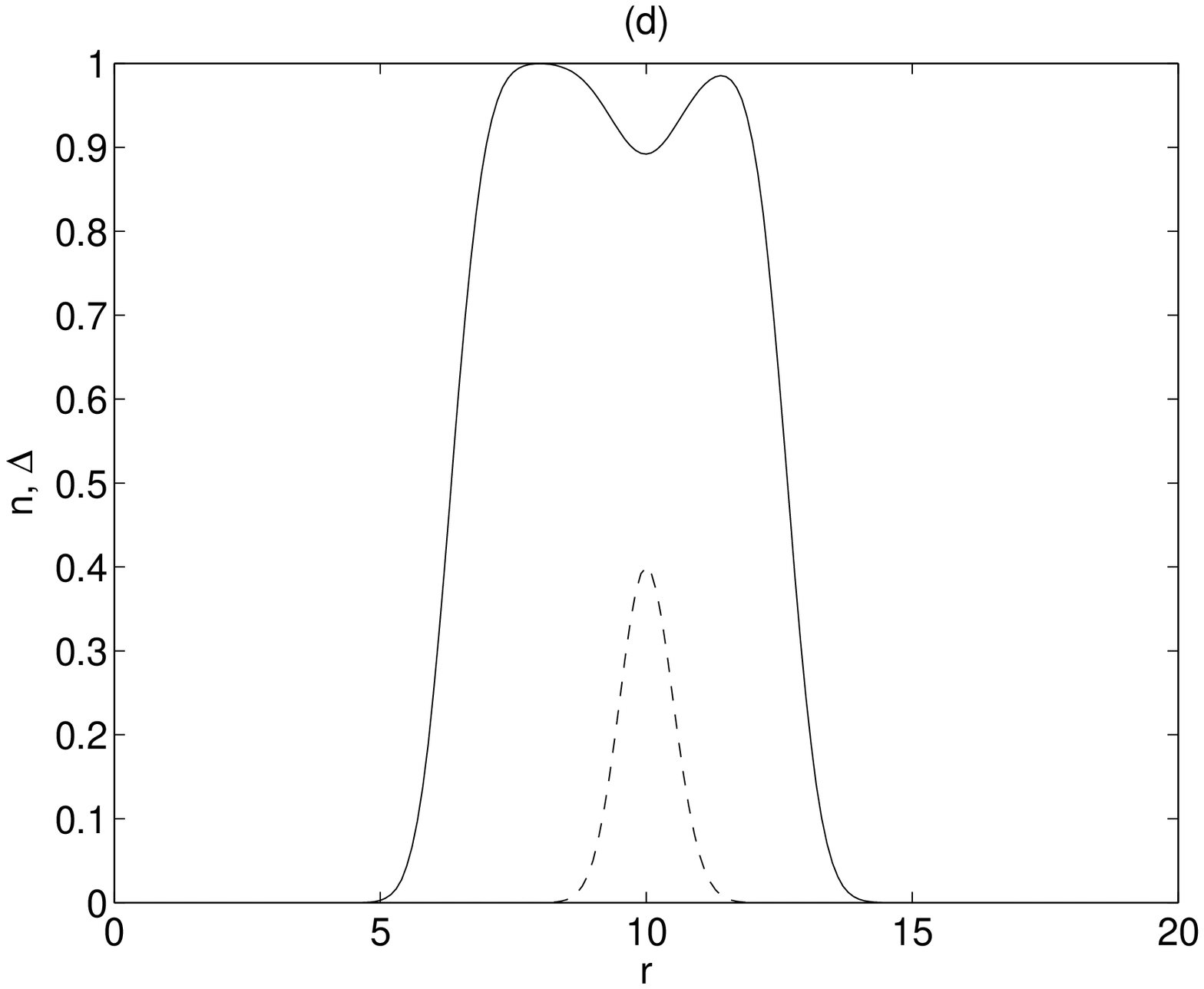}
\caption{\label{fig:qpspectrum}
(a),(c): Full line: Quasi-particle energies $E_{\kappa}$.
Dashed: $\kappa$-dependent gap function $\Delta_{\kappa}$. 
Dotted: summand $S_{\kappa}$ in the self-consistency sum. The 
gap function is magnified by a factor $10^8$ in (a). 
Horizontal line: chemical potential $\mu$.
(b), (d): Spatial distributions of density (full line) and gap 
function $\Delta(r)$ (dashed). The gap function is scaled to have 
a peak value of 0.4. Parameters are chosen 
as $m_0=100$, $\kappa_F=60$, and $\alpha=0.0001$. 
(a-b): $|g|=0.3$, 
(c-d): $|g|=3.0$. 
}
\end{figure}

The quasi-particle spectrum differs qualitatively between the 
different parameter regimes. This is illustrated in Fig.\ 
\ref{fig:qpspectrum}. 
When $|g|\lesssim g_c$, one recognizes 
the textbook situation: pairing takes place mostly at the Fermi 
surface and the spectrum is equal to the absolute value 
of the single-particle 
spectrum except close to the Fermi surface, where a gap opens. 
On the scale of Fig.\ \ref{fig:qpspectrum}a, the gap is too small 
to be visible. The plot also includes the $\kappa$-dependent gap function 
$\Delta_{\kappa}$ and the summand in the self-consistency 
equation, Eq.\ (\ref{finalself}), $S_{\kappa}=V_{\kappa}^2/E_{\kappa}$. It is 
seen that an appreciable contribution to the sum comes from terms 
close to the Fermi surface. If $|g|$ were even smaller, the sum 
would be entirely exhausted by the terms close to the Fermi surface, 
but as was found above, the BCS solution would be altered by 
effects associated with the discreteness of the spectrum.
For stronger coupling, $|g|> g_c$, 
the pairing happens predominantly close to $\kappa=0$, where the 
interaction is at its maximum, as seen in Figs.\ \ref{fig:qpspectrum}c-d. 
The quasi-particle spectrum is affected 
in this region, but at larger $\kappa$ it resembles the usual BCS spectrum.

The spatial profile of the gap function is completely 
insensitive to the profile in $\kappa$-space of the summand as long as 
the population is confined to the LLL. 
It is given by the self-consistency equation in 
configuration space, Eq.\ (\ref{selfconsistency}),
\beq
\Delta(r) 
= |g|\sum_{\kappa} \phi_{m_0+{\kappa}}(r) \phi_{m_0-{\kappa}}(r)
\frac{|\Delta_{\kappa}|}{2E_{\kappa}} 
= |g|\sum_{\kappa}\frac{|\Delta_{\kappa}|}{2E_{\kappa}} \frac1{\pi\sqrt{(m_0+{\kappa})!(m_0-{\kappa})!}}
r^{2m_0}e^{-r^2/2}.
\enq
Thus, the gap function is a constant times the $2m_0$'th radial 
harmonic-oscillator eigenfunction. At the same time, the density 
is much broader as a function of $r$, and it does 
depend on the coupling. The density is 
\beq
n(r) = 2\sum_{\kappa} |\phi_{m_0-{\kappa}}(r)|^2 |v_{\kappa}|^2,
\enq
with $|v_{\kappa}|^2= (1-\xi_{\kappa}/E_{\kappa})/2$. 
It is a peculiar feature of the 
LLL states that each eigenfunction $\phi_{m_0-{\kappa}}(r)$ is confined to 
a very narrow, shell-like region in configuration space, centered 
at $r=(m_0-{\kappa})^{1/2}$. As a 
result, the distribution of particles in $\kappa$-space is mirrored by the 
spatial density profile, giving direct observational access 
to the distribution of particles in angular momentum space. 
For weak coupling, all states below 
the Fermi surface contribute, just as in ordinary BCS theory. The 
spatial profile as a function of the radial coordinate is a 
flattened, almost square profile, nicely mirroring the filled Fermi sea. 
As a result, only the interior of the toroidal density distribution 
is superfluid, while the outer regions are normal.
For stronger coupling, the Bogoliubov 
amplitude $v_{\kappa}$ is depleted around $\kappa=0$ 
and the density profile has 
a dip in the region where the gap function is at its largest.
Mathematically, it is easy to see why the density is depressed at 
the trap bottom when the coupling is strong: the Bogoliubov 
amplitude $v_{\kappa}$ is equal to half its peak value when 
$E_{\kappa}=\sqrt{\Delta_{\kappa}^2+\xi_{\kappa}} \gg \xi_{\kappa}$. 
Physically, 
it is the strong interactions that alter the occupation of the 
single-particle levels. 
As a result, one obtains a spatial region of strong Cooper pairing 
that shows up as a density depression. In the next section, we shall 
examine the conditions for this effect to persist when higher 
Landau levels are included.

\section{NLLL}
\label{sec:nlll}
In order to approach slightly more realistic territory in parameter 
space, let us now study what happens when the next lowest Landau 
level (NLLL) is brought into the calculation. 
Inspection of Eq.\ (\ref{ineq1}) reveals that 
this means allowing the product $a N^2/16$ to be slightly 
larger than unity. This system is still not easily realizable in 
experiment, but it is a first correction to the even more difficult 
LLL case studied above. The new Landau level introduces a new parameter 
into the system, namely, the energy difference between Landau levels, 
which is equal to $2\omega=2$ in dimensionless units. Thus, the balance 
between three main parameters determine the physics of the system as well 
as the validity of the truncation of the basis. The three are the chemical 
potential $\mu$, the coupling strength $g$, and the energy difference 
$2\omega$. In addition, the balance between the Fermi level $\kappa_F$ and 
the width of the coupling, $\sqrt{m_0}$, plays a role in determining the 
physics.

The single-particle eigenfunctions in the NLLL are 
\beq
\phi_{m,1}(r) = \sqrt{\frac{m+1}{\pi m!}}\left(1-\frac{r^2}{m+1}\right)
(r e^{i\theta})^m e^{-r^2/2}.
\enq
The perturbative expression for the energies in the quartic trap are
(see Fig.\ \ref{fig:levels}) 
\beq
\epsilon_{m,1} = 3\omega -(\Omega-\omega)m + 
\frac{a}{2}(m+2)(m+4).
\enq
The position of the energy minimum of the NLLL coincides with that 
of the LLL, $m_0$, to within $3/2$ quanta. That shift can safely be 
neglected here.
On the other hand, when many Landau levels are brought into the 
calculation, the resulting asymmetry of the single-particle energy spectrum 
is likely to break the cylindrical symmetry such that the order 
parameter becomes a superposition of several angular-momentum 
states. The resulting profile is expected to be an annulus 
pierced by a ring of singly quantized vortices \cite{jackson2004}. 

Now consider the coupling matrix elements. They are 
\bea
V^{0001}_{\kappa,q} &=& \langle (m_0+\kappa,0),(m_0-\kappa,0)|
(m_0+q,0),(m_0-q,1)\rangle  
\nonumber\\
&=& \frac{(2m_0)! (1-2q)
}{2\pi 2^{2m_0}\sqrt{(m_0+q+1)!(m_0-q+1)!(m_0+\kappa)!(m_0-\kappa)!}}
\nonumber\\
&\approx & 
\frac{q-1}{\pi^{3/2}m_0}e^{-(\kappa^2+q^2)/(2m_0)} + {\mathcal O}(m_0^{-2}),
\\
V^{0011}_{\kappa,q} 
&=& \frac{1}{4\pi} \frac{(2m_0)! (m_0+1-2 q^2)
}{2^{2m_0}\sqrt{(m_0+q+1)!(m_0-q+1)!(m_0+\kappa)!(m_0-\kappa)!}},
\nonumber\\
&\approx & 
\frac{1 - 2q^2/m_0}{4\pi\sqrt{\pi m_0}}e^{-(\kappa^2+q^2)/(2m_0)},
\\
V^{0101}_{\kappa,q} 
&=& \frac{(2m_0)! (1+m_0-\kappa-q+4\kappa q)
}{8\pi 2^{2m_0}\sqrt{(m_0+q)!(m_0-q+1)!(m_0+\kappa)!(m_0-\kappa+1)!}}
\nonumber\\
&\approx & 
\frac{1}{4\pi\sqrt{\pi m_0}}e^{-(\kappa^2+q^2)/(2m_0)}
\left(1+\frac{4\kappa q-q-\kappa}{2m_0}\right),
\\
V^{1101}_{\kappa,q} 
&=& \frac{(2m_0)! (1+m_0-2q-2m_0q-2\kappa^2+4\kappa^2q)
}{2\pi 2^{2m_0}\sqrt{(m_0+q)!(m_0-q+1)!(m_0+\kappa+1)!(m_0-\kappa+1)!}}
\nonumber\\
&\approx & 
\frac{1-2q}{2\pi^{3/2}m_0}e^{-(\kappa^2+q^2)/(2m_0)},
\\
V^{1111}_{\kappa,q} &=& \frac{1}{8\pi} \frac{(2m_0)! [2+3m_0^2-2(\kappa^2+q^2)
+m_0(5-2(\kappa^2+q^2)) +4\kappa^2q^2]
}{2^{2m_0}\sqrt{(m_0+q+1)!(m_0-q+1)!(m_0+\kappa+1)!(m_0-\kappa+1)!}}
\nonumber\\
&\approx & \frac{3}{8\pi\sqrt{\pi m_0}}
(1-\frac{1+2(\kappa^2+q^2)}{3m_0})
e^{-(\kappa^2+q^2)/(2m_0)}.
\ena
The other matrix elements can be obtained from these by a sign change, 
e.~g.\ $V^{0010}_{\kappa,q}=V^{0001}_{\kappa,-q}$.
It is seen that these new matrix elements are not separable in the way 
that simplified the calculations within the LLL.
Moreover, the matrix elements $V^{0001}$ and $V^{1101}$ are 
smaller than the others by a factor $m_0^{1/2}$; 
however, that does not mean that  the inter-level coupling 
cannot be neglected, since the matrix elements $V^{0011}$ and 
$V^{0101}$ are still as large as the intra-level matrix elements.

Employing the cylindrically symmetric assumption for $\Delta(r)$ 
(keeping in mind that this may change when 
more Landau levels are brought into the problem), 
the ansatz for the Bogoliubov amplitudes has to be labeled by 
an angular momentum quantum number $\kappa$ and a branch index 
$j=0,1$ (enumerating the two solutions that exist for every $\kappa$),
\bea
u_{\kappa j}(r) = u_{\kappa j0} \phi_{m_0+\kappa,0}(r) + 
u_{\kappa j1}\phi_{m_0+\kappa,1}(r),
\nonumber\\
v_{\kappa j}(r) = v_{\kappa j0} \phi_{m_0-\kappa,0}^*(r) + v_{\kappa j1}\phi^*_{m_0-\kappa,1}(r).
\ena
The Bogoliubov equation for the $(k,j)$'th mode is
\beq
\left(\begin{array}{llll} 
\xi_{\kappa 0} & 0 & \Delta_{\kappa}^{00} & \Delta_{\kappa}^{01} \\
0 & \xi_{\kappa 1} & \Delta_{\kappa}^{10} & \Delta_{\kappa}^{11} \\
{\Delta_{\kappa}^{00}}^* & {\Delta_{\kappa}^{10}}^* &-\xi_{{\kappa}0} & 0 \\ 
{\Delta_{\kappa}^{01}}^* & {\Delta_{\kappa}^{11}}^* & 0 & -\xi_{{\kappa}1}  
\end{array}\right) 
\left(\begin{array}{l}
u_{{\kappa}j0}\\u_{{\kappa}j1}\\v_{{\kappa}j0}\\v_{{\kappa}j1}
\end{array}\right) =
E_{{\kappa}j} \left(\begin{array}{l}
u_{{\kappa}j0}\\u_{{\kappa}j1}\\v_{{\kappa}j0}\\v_{{\kappa}j1}
\end{array}\right).
\enq
We have defined $\xi_{{\kappa}n}=\epsilon_{{\kappa}n}-\mu$ and the matrix elements
\beq
\Delta_{\kappa}^{nn'} = \int d^2r \Delta(r) \phi_{m_0+{\kappa},n}^* \phi_{m_0-{\kappa},n'}^*.
\enq
The self-consistency equation is in general
\beq
\Delta(r) = |g|\sum_{{\kappa}j} u_{{\kappa}j}(r)v_{{\kappa}j}(r)^*,
\enq
and integrating one obtains
\beq
\Delta^{n_1n_2}_{\kappa} = |g|\sum_{{\kappa}'j,n_3,n_4} 
u_{j{\kappa}'n_3} v^*_{j{\kappa}'n_4} V^{n_1n_2n_3n_4}_{{\kappa}{\kappa}'}.
\enq

\begin{figure}
\includegraphics[width=0.48\columnwidth]{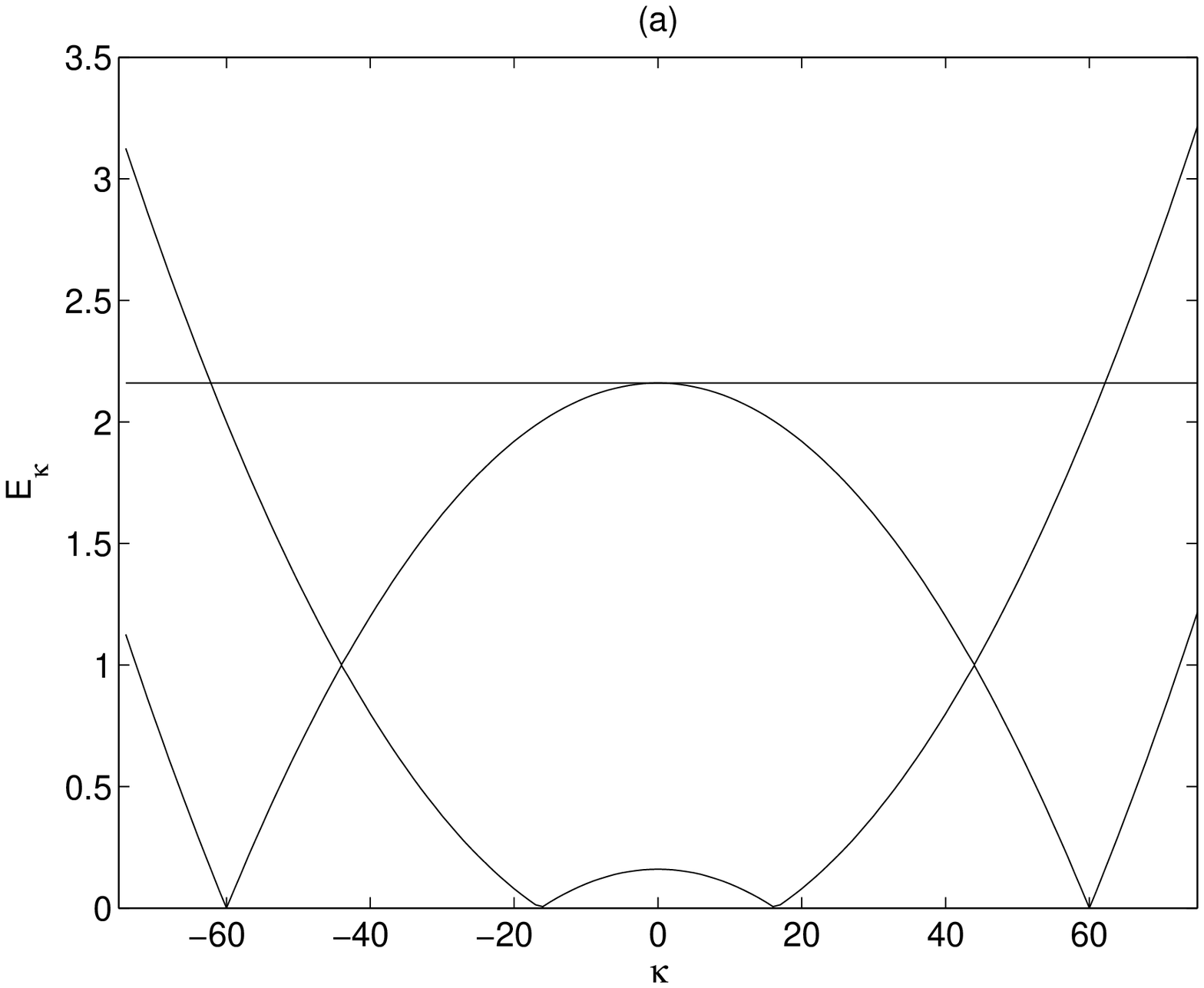}
\includegraphics[width=0.48\columnwidth]{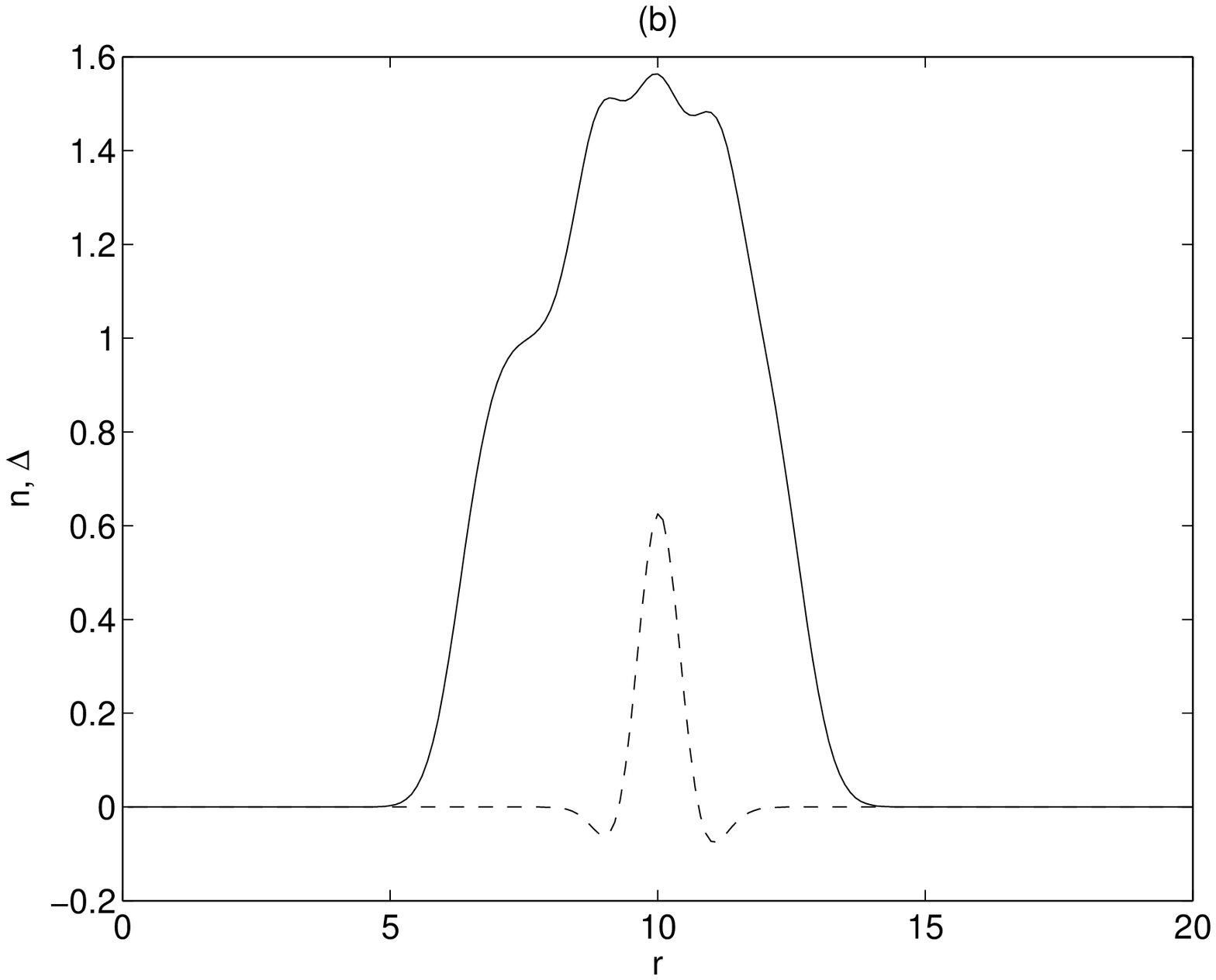}
\includegraphics[width=0.48\columnwidth]{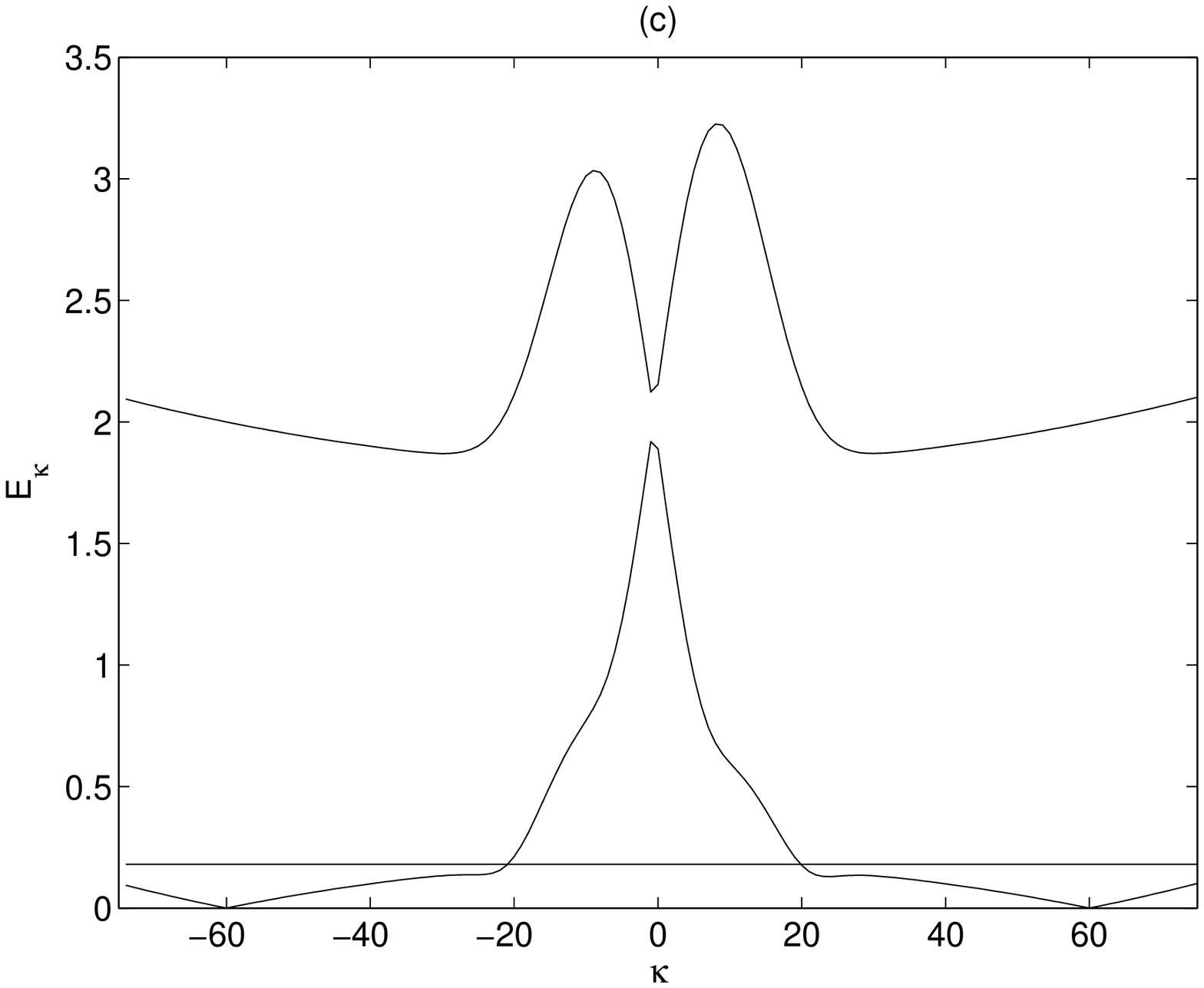}
\includegraphics[width=0.48\columnwidth]{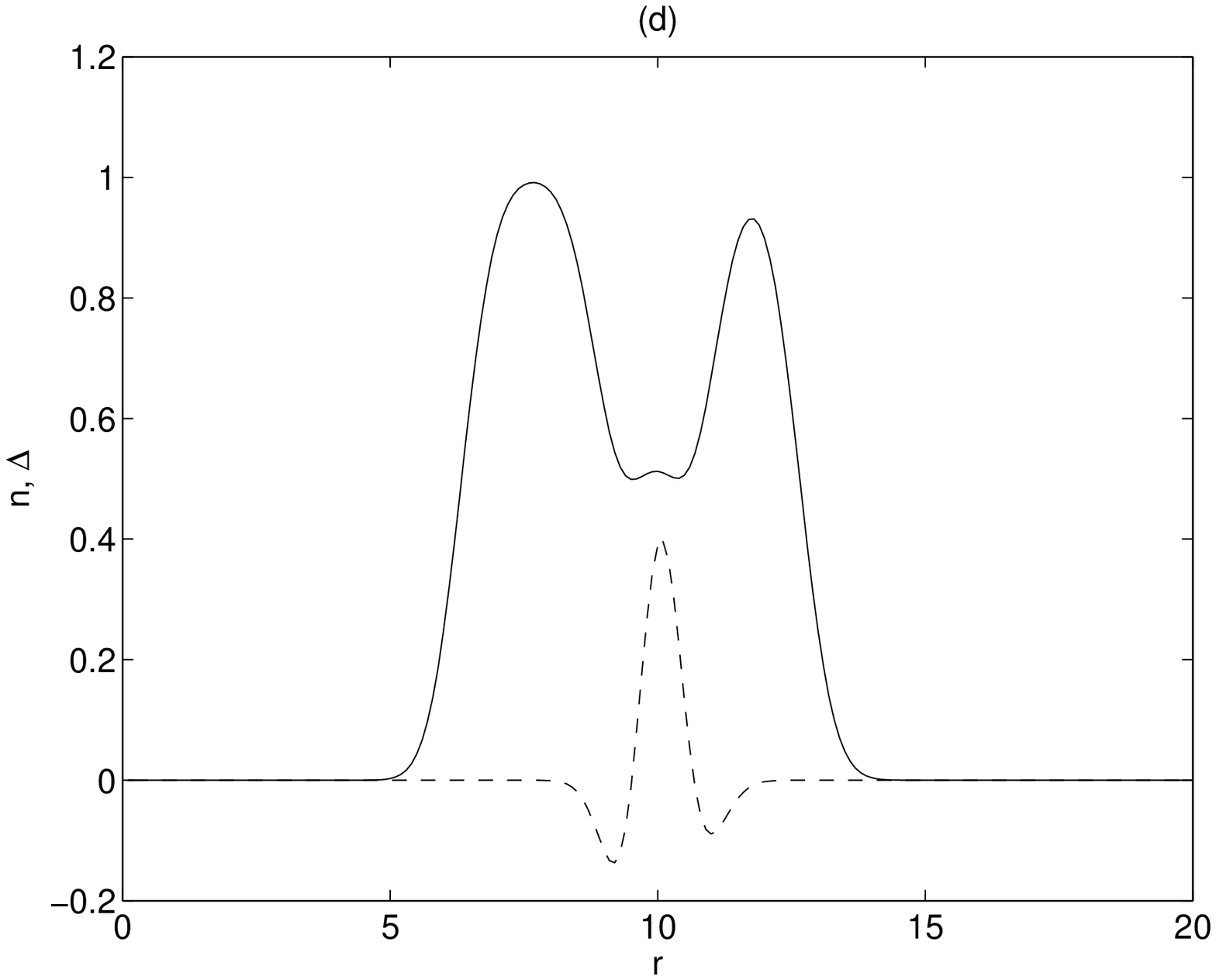}
\includegraphics[width=0.48\columnwidth]{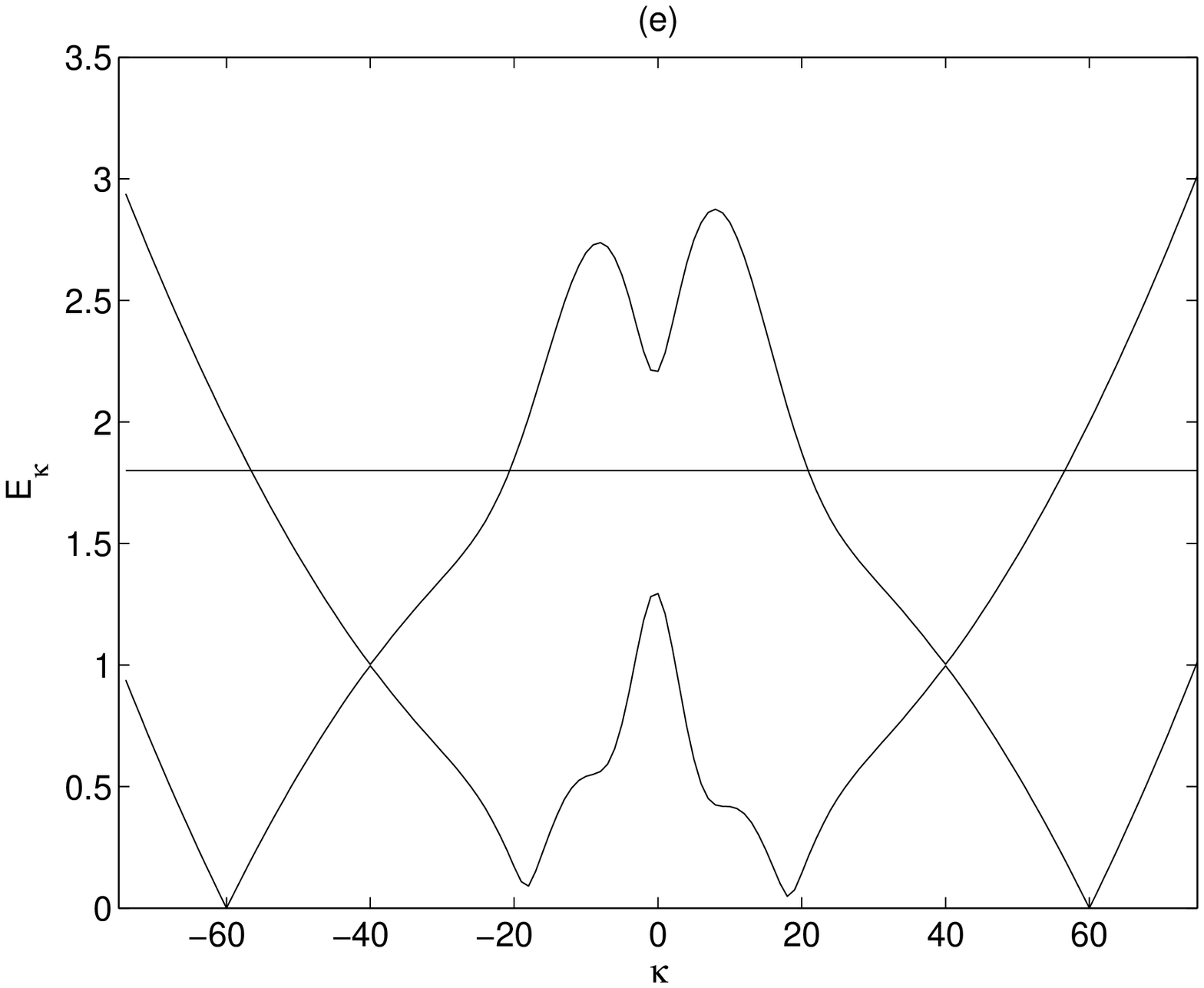}
\includegraphics[width=0.48\columnwidth]{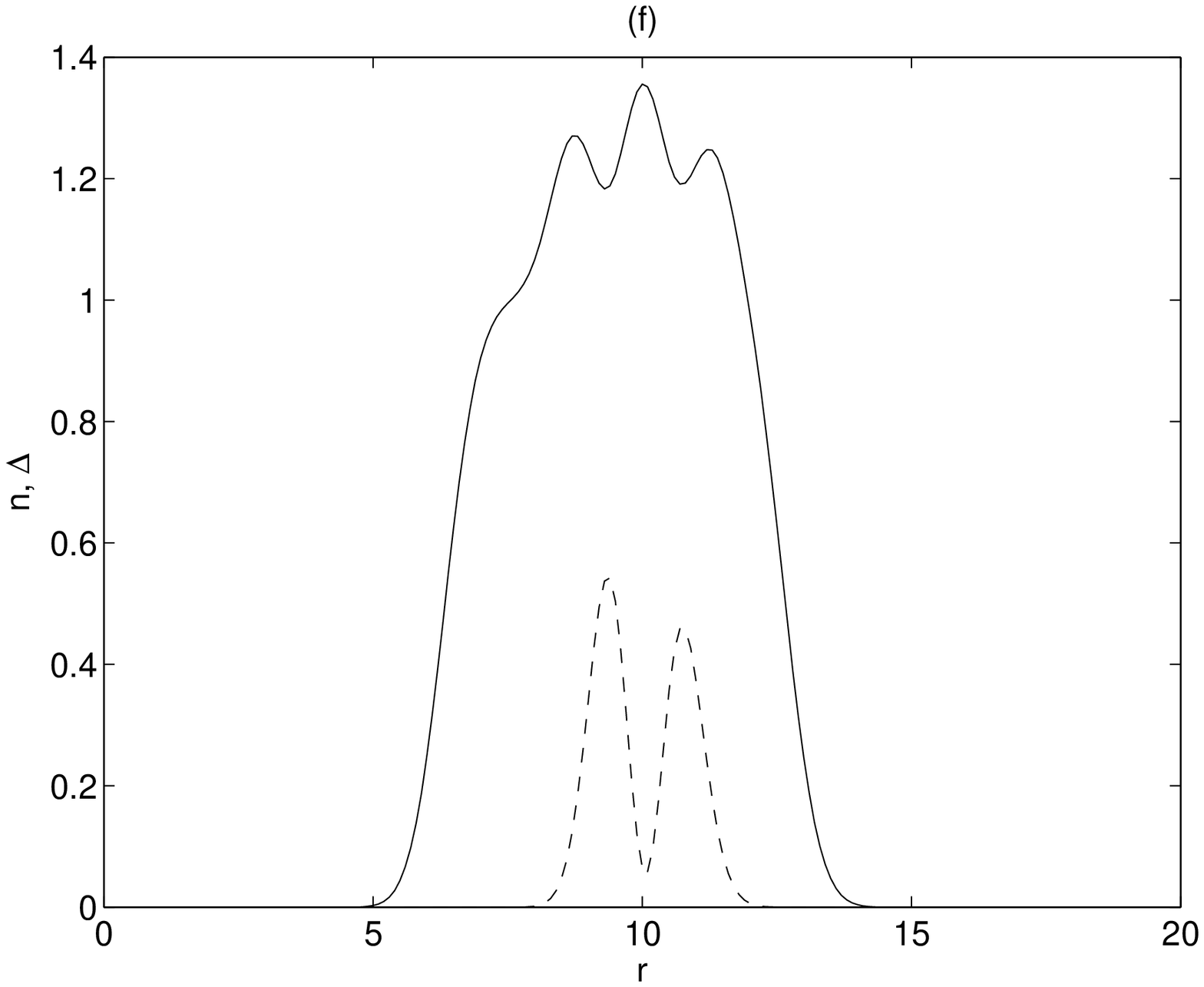}
\caption{\label{fig:nlll}
Energy spectra and density profiles calculated using two
Landau levels (NLLL approximation). 
(a),(c),(e): Quasi-particle energies $E_{\kappa}$.
Horizontal line: chemical potential $\mu$.
(b), (d), (f): Spatial distributions of density (full line) and gap 
function $\Delta(r)$ (dashed). The gap function is scaled to have 
a peak value of 0.4 times the maximum density. Parameters are chosen 
as $m_0=100$, and ${\kappa}_F=60$. 
(a-b): $|g|=0.4$ and $\alpha=0.0012$, 
(c-d): $|g|=8.0$ and $\alpha=0.0001$, (e-f): $|g|=8.0$ 
and $\alpha=0.001$. 
}
\end{figure}

The coupled equations are solved numerically. 
Figure \ref{fig:nlll} displays a few illustrative 
cases. 
It has been checked in trial calculations that the inclusion of more 
Landau levels does not alter the density profiles shown in this paper, 
and only marginally distorts the quasi-particle energy curves. 
Basically, there are four parameter regimes of interest. When 
$\mu \ll 1$ and $|g| \ll 1$, the NLLL is not occupied at all. 
This case was studied in the previous section. 
When $\mu \sim 1$ and $|g| \ll 1$, 
the Fermi surface cuts through both energy bands, but the 
coupling is not too strong, as in Figs.\ \ref{fig:nlll}a-b. There 
is a gap at the bottom of each of the two bands. The 
NLLL has a larger gap because the coupling is stronger for smaller ${\kappa}$. 
The associated density profile is expected to be bimodal, with an 
enhanced density in the interior of the toroidal cloud where 
two Landau levels are populated (cf.\ \cite{ho2000}). 
Because the Fermi level has been limited to 
${\kappa}_F=60$ in the numerical calculation, this ``wedding cake'' 
structure does not 
come out as clearly in Fig.\ \ref{fig:nlll}b as it would in a bigger 
system.
Figs.\ \ref{fig:nlll}c-d illustrates the case where the chemical 
potential is small, but the coupling is 
comparable to the separation between the energy bands; 
$\mu \ll 1$ and $|g| \sim 1$. 
The Fermi level lies far below the NLLL, but because of the strong 
coupling, both levels are deformed. The resulting density profile 
is still depleted in the region of strong Cooper pairing, 
because the contribution from the NLLL is modest. For stronger 
interactions, more Landau levels are 
coupled and the NLLL approximation would fail. Note that 
the effect of the strong coupling on the quasi-particle spectrum is much 
more pronounced than the effect on the density.
Finally, in Fig.\ \ref{fig:nlll}e-f, both $\mu$ and $|g|$ are 
comparable to the inter-level gap. The energy levels are 
heavily distorted by the strong coupling. The density near 
$r=m_0^{1/2}$ is no longer depleted, because the NLLL states 
now occupy that regime. 
The gap function takes on a slightly nontrivial 
shape in configuration space; this is again an effect 
of the relatively few single-particle levels involved in the 
calculation. Trial calculations indicate that when the number of 
states is increased, the gap function acquires a smoother shape, but 
the contribution from the radially excited states broadens its profile 
compared to the LLL case. 

Extrapolating from the results presented here, one may observe 
that the density is affected by two competing effects: as the Fermi 
surface includes more and more Landau levels, a 
``wedding cake'' structure is expected in the density for weak 
interactions, but at the same time, strong interactions will deplete 
the interior of the toroidal density distribution.

\section{Conclusion and outlook}
\label{sec:conclusions}
This paper has shown that a Fermi superfluid trapped in an anharmonic 
potential can develop a giant vortex if it is rotated at high enough 
angular velocity, since the single-particle energy spectrum will 
then have its global minimum at a finite angular momentum. 
A giant-vortex state 
has been constructed in the basis of the lowest radially excited 
states, the lowest Landau level (LLL). Although attaining the LLL 
regime experimentally puts severe constraints on 
precision, the theoretical construction of such a state demonstrates that 
there exist parameter regimes in anharmonic traps for which a Fermi 
gas develops a giant vortex. As long as only the LLL is occupied, 
the density and gap function are cylindrically symmetric 
and form a toroidal profile. The superfluid region is confined to the 
interior of the torus, while the edges are normal. Since 
Cooper pairing is strongest in 
the center of the toroidal profile, the density 
is depleted in that region. The effect on the density appears, 
however, to be exclusive to the LLL. As more Landau levels are 
occupied, singly quantized vortices may form within the torus.

In order to have only the lowest Landau level populated in the 
quadratic plus quartic trap, the anharmonicity and rotation 
frequency have to be extremely fine tuned. In addition, any 
mechanical stirring method \cite{madison2000} will bring in 
corrections to the anharmonic trap that may be hard to control. 
It is therefore probably better to control the angular 
momentum of the gas by first stirring it in a harmonic potential 
and subsequently applying the anharmonicity. As an alternative to 
an extremely small quartic trapping term, an 
optical hard-wall potential at a suitably large distance from 
the trap center may be a more practical choice. It is possible that 
such a setup, with extensive amounts of fine tuning, may be a more 
practical route towards accessing the lowest Landau level.

Nevertheless, this study should first and foremost be seen as 
a demonstration by construction 
that trapped Fermi superfluids can sustain giant vortices.
Several important issues 
remain to be studied, notably the size and shape of the allowed 
parameter regime for giant vortices; 
the physics in the BEC-BCS crossover regime; and 
whether giant vortices can ever be the favorable rotational 
configuration in purely harmonic traps. We 
hope to address these issues in future work.

\begin{acknowledgments}
The author would like to thank Jani Martikainen and 
Andrei Shelankov for helpful discussions.
This work was financed by the Swedish Research Council.
\end{acknowledgments}



\begin{thebibliography}{99}
\bibitem{ll} E.~M.\ Lifshitz and L.~P.\ Pitaevskii, 
{\it Statistical Physics Part 2} (Pergamon Press, Oxford, 1980).
\bibitem{pethick2001} C.~J.\ Pethick and H.\ Smith, {\it Bose-Einstein
  Condensation in Dilute Gases}
  (Cambridge University Press, Cambridge, 2001).
\bibitem{madison2000} K. W. Madison, F. Chevy, W. Wohlleben, and J. Dalibard,
  Phys.\ Rev.\ Lett.\ {\bf 84}, 806 (2000).
\bibitem{ketterle2001} C.\ Raman, J.~R.\ Abo-Shaeer, 
  J.~M.\ Vogels, K.\ Xu, and W.\ Ketterle,  Phys.\ Rev.\ Lett.\ 
  {\bf 87}, 210402 (2001).
\bibitem{ketterle2005} M.~W.\ Zwierlein, J.~R.\ Abo-Shaeer, 
  A.\ Schirotzek, C.~H.\ Schunck, and W.\ Ketterle, 
  Nature {\bf 435}, 1047 (2005).
\bibitem{deo1997} P.\ Singha Deo, V.~A.\ Schweigert, and 
  F.~M.\ Peeters, Phys.\ Rev.\ Lett.\ {\bf 79}, 4653 (1997).
\bibitem{butts1999} D.~A.\ Butts and D.~S.\ Rokhsar, Nature {\bf 397}, 
  327 (1999).
\bibitem{kavoulakis2000} G. M. Kavoulakis, B. Mottelson, and C. J. Pethick,
  Phys. Rev. A {\bf 62}, 063605 (2000).
\bibitem{lundh2002} Emil Lundh, Phys.\ Rev.\ A {\bf 65}, 043604 (2002).
\bibitem{kasamatsu2002} K.\ Kasamatsu, M.\ Tsubota, and M.\ Ueda,
  Phys.\ Rev.\ A {\bf 66}, 053606 (2002).
\bibitem{fischer2003} Uwe R. Fischer and Gordon Baym,
  Phys.\ Rev.\ Lett.\ {\bf 90}, 140402 (2003).
\bibitem{kavoulakis2003} G.~M.\ Kavoulakis and G.\ Baym, New J.\ 
  Phys.\ {\bf 5}, 51 (2003).
\bibitem{aftalion2003} A.\ Aftalion and I.\ Danaila, 
  Phys.\ Rev.\ A {\bf 69}, 033608 (2004).
\bibitem{jackson2004} A.~D.\ Jackson, G.~M.\ Kavoulakis, and E.\ Lundh,
  Phys.\ Rev.\ A {\bf 69}, 053619 (2004). 
\bibitem{fetter2005} A.~L.\ Fetter, B.\ Jackson, and S.\ Stringari, 
  Phys.\ Rev.\ A {\bf 71}, 013605 (2005).
\bibitem{fu2006} H.\ Fu and E.\ Zaremba, Phys. Rev. A {\bf 73}, 
  013614 (2006).
\bibitem{bargi2006} S.\ Bargi, G.~M.\ Kavoulakis, and S.~M.\ Reimann, 
  Phys. Rev. A {\bf 73}, 033613 (2006).
\bibitem{lundh2006} E.\ Lundh and H.~M.\ Nilsen, e-print 
  cond-mat/0608532 (2006).
\bibitem{toreblad2004} M.\ Toreblad, M.\ Borgh, M.\ Koskinen, 
  M.\ Manninen, and S.~M.\ Reimann,
  Phys.\ Rev.\ Lett.\ {\bf 93}, 090407 (2004).
\bibitem{sensarma2006} R.\ Sensarma, M.\ Randeria, and T.-L.\ Ho, 
  Phys.\ Rev.\ Lett.\ {\bf 96}, 090403 (2006).
\bibitem{delft}  F.\ Braun and J.\ von Delft,
  Phys.\ Rev.\ Lett.\ {\bf 81}, 4712 (1998).
\bibitem{ho2000} T.-L.\ Ho and C.~V.\ Ciobanu, Phys.\ Rev.\ Lett.\ 
  {\bf 85}, 4648 (2000).
\end{thebibliography}
\end{document}